# Classification of Neurodevelopmental Age in Normal Infants Using 3D-CNN based on Brain MRI

Mahdieh Shabanian
*Department of Biomedical Engineering*
University of Tennessee Health Science Center & University of Memphis
Memphis, USA

Eugene C. Eckstein
*Department of Biomedical Engineering*
University of Tennessee Health Science Center & University of Memphis
Memphis, USA

Hao Chen
*Department of Pharmacology*
University of Tennessee Health Science Center
Memphis, USA

John P. DeVincenzo
*Department of Pediatrics*
University of Tennessee Health Science Center
Le Bonheur Children's Hospital
Memphis, USA

*Abstract*— Human brain development is rapid during infancy and early childhood. Many disease processes impair this development. Therefore, brain developmental age estimation (BDAE) is essential for all diseases affecting cognitive development. Brain magnetic resonance imaging (MRI) of infants shows brain growth and morphologic patterns during childhood. Therefore, we can estimate the developmental age from brain images. However, MRI analysis is time consuming because each scan contains millions of data points (voxels). We investigated the three-dimensional convolutional neural network (3D CNN), a deep learning algorithm, to rapidly classify neurodevelopmental age with high accuracy based on MRIs. MRIs from normal newborns were obtained from the National Institute of Mental Health (NIMH) Data Archive. Age categories of pediatric MRIs were 3 wks ± 1 wk, 1 yr ± 2 wks, and 3 yrs ± 4 wks. We trained a BDAE method using T1, T2, and proton density (PD) images from MRI scans of 112 individuals using 3D CNN. Compared with the known age, our method has a sensitivity of 99% and specificity of 98.3%. Moreover, our 3D CNN model has better performance in neurodevelopmental age estimation than does 2D CNN.

## I. Introduction

Rapid human brain development occurs during infancy and early childhood. Many disease processes impair this development. Brain developmental age estimation (BDAE) is, therefore, an essential component of all diseases impacting cognitive development. Brain magnetic resonance imaging (MRI) of infants shows brain growth and morphologic patterns during childhood. This technique also provides qualitative and quantitative assessments of neural development to elucidate early brain growth, morphology, and pattern changes during infancy. An infant's brain develops rapidly during the first few months and years of life. The normative pattern of infant brain development, as assessed by MRI, is complex and has been incompletely determined. Neuroimaging-derived brain age is crucial in assessing pediatric brain diseases. Healthy clinical sample brain-MRI data can be used to train a deep-learning program for healthy and related infant brains. Such a standard could be permit comparisons of normal brain development with that in many infant diseases process including prematurity [1], Hypoxic Ischemic Encephalopathy (HIE) of the newborn [2], congenital Cytomegalovirus (cCMV) infection [3], and Bacterial Meningitis, and Herpes Simplex Virus encephalitis (HSVE) [4].

Newborn with these and other disorders are at increased risk of the severe long-term motor deficit, and cognitive delay. Early detection of CNS infection can help mitigate morbidity. Statistical analysis of brain magnetic resonance (MR) images reveals a specific pattern in brain morphological changes during brain development and maturation, specifically in the first three years of life; deviations from these patterns can be attributable to either lack of growth or abnormal growth. Volumetric changes of brain tissues, like white matter (WM), gray matter (GM), and cerebrospinal fluid (CSF), are illustrated during the normal maturities process [5, 6, 7, 8]. We hypothesized that deep learning approaches could be used to classify normal age specific MRI brain development. Such a normative classification could then be applied as a morphological biomarker for early identification and diagnosis of age-related brain disorders in infants.

One major challenge is classifying the neonatal brain (less than 1 month) due to lower signal to noise ratio and the small size of their brain. Additionally, water content is high in the neonatal brain, and the grey matter has yet to fully develop. During the first year, water content rapidly decreases, and a conventional interpretation of grey scales provides a correct assessment for development. Between 2 to 3 years, the brain myelinates and becomes more similar to the MRI of an adult brain. Rapid myelination also occurs in the brain in the first three years. Additionally, reliable cognitive testing can be performed at the earliest age of one year. Thus, this testing provides important functional validation of normative MR imaging at this age. Therefore, Pediatricians and researchers are interested in younger age cohorts (ie, neonate, one year, and 3 years) to diagnose any subtle structural abnormalities at risk children.

The dearth of available digital tools for brain MRI analysis has hindered quantitative studies of infant brain development. However, recent progress in computer vision through deep neural networks is moving to applications in medical image analysis [6, 9, 10]. Lengthy post-processing represents another limitation in clinical applications. Analyzing MRI scans after acquisition can take hours or days because each scan contains millions of voxels. By contrast, clinical decisions are needed in minutes or less in some severe diseases in infants and require highly specialized expertise developed over years of practice. Implementing intelligent algorithms to automate medical image analysis and increase the accuracy of image correlations with an infant's disease state could save time and money for radiologists, hospitals, parents, and infants. In recent years, artificial neural networks and machine learning methods have become popular and are used widely in various classification and clustering tasks [11, 12]. Moreover, deep learning (DL) algorithms offer an automated approach to statistical modeling in neuroimaging like the BDAE proposed here. Deep learning provides advantages for high-dimensional estimation tasks that may enable learning of both latent relationships and physiologically relevant representations [13].

## II. Deep learning

Deep learning algorithms, specifically convolutional networks, have quickly become the best choice for analyzing medical images. These algorithms represent a shift from systems designed by humans to systems trained by computers using input data to extract feature vectors automatically [5]. DL techniques provide more accurate results while lessening pre- and post-processing tasks during MRI analysis [5, 6].

DL algorithms (networks) have many layers of artificial neurons that transform input data (e.g., MR images) to outputs (e.g., neurodevelopmental age classification). By learning from selected data, these algorithms will increasingly include higher-level features to reduce an image to its key features, thus enabling easier classification. Currently, convolutional neural networks (CNNs) algorithms are popular models for image classification, video analysis, facial recognition [14, 15]. CNNs contain many layers that transform their input with small convolution filters automatically. The first CNN was designed in the late 1970s [16]. Lo et al. [17] used CNNs for medical image analysis in 1993. They saw their first successful real-world application in LeNet [18]. Despite these initial successes, after several years of interruptions in developing a new model in DL, Fei-Fei Li started ruminating on an idea called ImageNet in 2006 [19]. The annual ImageNet competition is known as the ImageNet Large Scale Visual Recognition Challenge (ILSVRC). Krizhevsky et al. [20] Introduced a new model in the ILSVRC in December 2012. Their proposed deep convolutional neural net, called AlexNet, won that competition. In the following years, more progress has been made using related but deeper architectures, by Russakovsky et al.[21]. Thus, deep convolutional networks have become the technique of choice in image processing [9]. In seven years, the winning accuracy in classifying objects in the dataset increased from 71.8% to 97.3% in CNN, surpassing human abilities, and conclusively proving that more accurate data lead to better decisions [21].

There are various types of CNNs being developed, and all have the potential to contribute to the speed and accuracy of image identification automatically. CNN is designed to learn high-level hierarchies of features automatically by backpropagation with multiple building blocks, such as convolution layers, max-pooling layers, and fully connected layers used for image classification tasks.

### A. 2D CNN Model

In 2D CNNs, convolutions are applied to 2D feature maps to compute features from spatial dimensions only. In fact, 2D CNNs take a 2D matrix as input, like image slices to segment or classify the images. Therefore, we do not have the information of adjacent slices. With this approach, losing information from interest regions of the brain is more likely in specific diseases because the brain is a 3D structure. Fig.1(a) shows the 2D filter (kernel) in 2D CNN to extract spatial-spectral features.

### B. 3D CNN Model

Architectures with volumetric (i.e., cube) convolutions have been successfully used in medical and video analysis [3]. From our review of the literature, 3D CNNs are appropriate deep methods for analyzing imaging data, a time-consuming process that currently requires expert analysis, especially for medical images such as Neuroimaging Informatics Technology Initiative (NIfTI) file formats of MRI [7, 8, 22, 23]. For the first time, Cole et al. [22], showed that 3D-CNN could accurately estimate developmental brain age from MRI data using healthy adult samples. Also, they demonstrated that 3D CNNs are more effective and less likely to miss regions of interest in medical images. Therefore, using 3D CNNs is often a superior approach, especially in cases where post-processing best practices have not been established [6].

3D convolution applies a 3D filter (kernel) to the dataset, and this filter moves in 3-directions (x, y, z) to calculate low-level feature representations (Fig.1(b)). The output shape is a 3D volume space such as a cube. 3D CNNs help to prevent loss of the region of interest in 3D and 2D medical images. In fact, CNNs apply a filter to the input to create a feature map that summarizes the presence of detected features in the input. Filters, such as line detectors, edges, curve lines can be handcrafted, but the innovation of CNNs is to learn the filters during training in the context of a specific classification problem.

### C. Hardware and Software

Training 3D and 2D CNNs require a graphic processing unit (GPU) with memory matched to the size of images to be analyzed. GPUs are a major hardware part that led to the remarkable rise of deep learning. Current GPUs typically increase speed by 10 to 50 times of CPU-based work. The success of AlexNet developed by Krizhevsky et al. is due to its architecture, the power of calculation on the GPU.

Several open-source DL software packages are available. These libraries provide efficient GPU implementations of deep neural networks, such as convolutions [6]. In 2015, Google Brain developed TensorFlow [24], which provides C++ and Python interfaces and is used by Google AI. Google's team developed Keras, a popular framework that provides the Python interface, in 2017.

## III. METHODS

### A. Dataset

The 317 MRI images of 112 infants for this study were obtained from the NIMH Data Archive (NDA). The NIH pediatric MRI images came from infants and young children ranging in age from 8 days to 3 years. All infants were screened for the presence of factors that might adversely affect brain development (e.g., preterm birth, physical growth delay, perinatal complications, learning disabilities, and psychiatric disorders of first-order family members). The infants were scanned with 1.5T MRI, while awake, or during natural sleep without sedation. This cohort age included: 2 weeks (8 to 35 days); 12 months (each ± 2-weeks) and 3 years (each ± 4-weeks). In this paper, we analyzed and classified these three clinically important and challenging age cohorts with respect to their brain MRI. Fig. 2 shows the number of subjects in each age cohort (24 subjects in newborn, 33 subjects in 1 year and 55 subjects in 3 years). Each subject had T1, T2, and PD except for 19 subjects that had only two of three MRI modalities in the dataset.

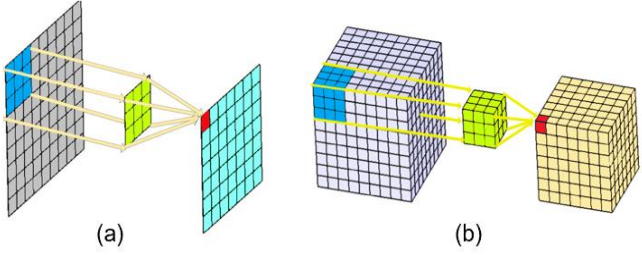

Fig. 1. Illustration of convolution: (a) illustration of 2D filter to extract spatial features; (b) illustration of a 3D filter to extract spatial-spectral features.

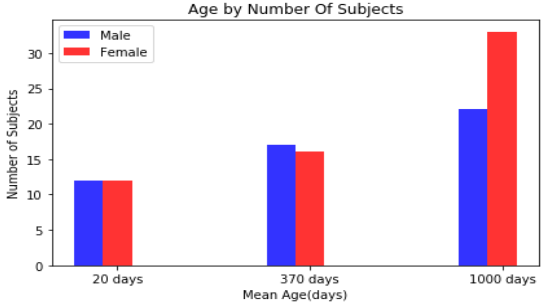

Fig. 2. Age distribution of the NIMH data archive.

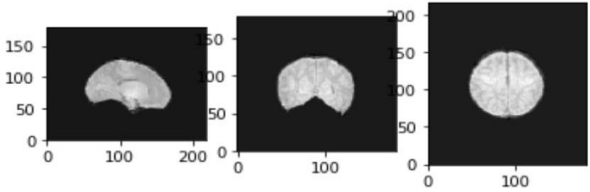

Fig. 3. Raw NIfTI MRI.

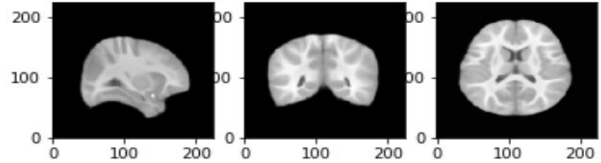

Fig. 4. The resized MRI.

All participants recruited were born full-term (> 37 weeks, 3 days), with equal representation of males and females. Income, race, and ethnicity were distributed in a demographically balanced sample to mirror proportions defined by the United States Census Bureau in 2000.

### B. Preprocessing

In this dataset, we had a different type of shape. The shapes of the scans became (192, 256, 48), (256, 256, 46), and (256, 256, 56). Fig.3 shows the specific slice of raw NIfTI image in each axial, sagittal, and coronal planes with different size after loading. The first significant step for pre-processing in DL is to resize all images to the same size in the axial, sagittal, and coronal planes, for example, to 200 pixels. Fig.4 shows the example of resized images to (200, 200, 200).

## IV. MEASURING METRICS

In the following, the statistical measures for evaluating the performance of classification methods are described. True positive (TP), False Positive (FP), True Negative (TN), and False Negative (FN) values are used for calculating performance measures [7]. TP denotes the number of instances that belong to a class and recognized accurately by the proposed modeling as the members of that class. FP denotes the number of instances that are wrongly recognized as the members of the class, but they actually belong to other classes. This number of instances truly detected as do not belong to a class, and FN is the number of instances that wrongly recognized as not belonging to a class. The accuracy, recall, precision, and F1 score of a model (BDAE system) are calculated by equations (1) through (4).

$$Accuracy = \frac{(TP+TN)}{(TP+TN+FP+FN)} \quad (1)$$

$$Recall = \frac{TP}{(TP+FN)} \quad (2)$$

$$Precision = \frac{TN}{(TP+FP)} \quad (3)$$

$$F1-score = \frac{(2*Precision*Recall)}{(Precision+Recall)} \quad (4)$$

The normalized confusion matrices show the percentage of numbers is correct, and the numbers of other cells show the error of the BDAE classification system.

## V. EXPERIMENTS AND RESULTS

In this study, two deep CNN architectures based on 2D CNN, and 3D CNN were used and evaluated for automatic neurodevelopmental age classification in three major age cohorts in infants by T1, T2 and PD images. We used the NVIDIA TITAN RTX GPU with Keras (latest release version 2.2.4) as DL frameworks. To reduce computational complexity all NIfTI MRIs applied to a gray and resized them to (80, 80, 80) as input data for 2D and 3D models. The training set and validation set were randomly selected as 80% and 20% of the dataset respectively. We used 253 MRI scans (47 newborns, 80 one year and 126 three years) for training and 64 MRI scans (11 newborns, 18 one year and 35 three years) for validation. In total, we used 37,950 slices of axial, sagittal and coronal for training and 22,400 slices for validation (Test). The proposed models were trained with the fusion of T1, T2 and PD in each age cohort.

### A. 2D CNN Model

An optimum set of hyperparameters is required of training a network to get the minimum differences between output prediction and given label MRI. Backpropagation algorithms are used commonly for training convolutional neural networks. Also, loss function and optimization algorithms play a key role in having a minimum loss. In our model, cross entropy and RMSprop were used respectively for the loss function and optimizer. The learning rate is another important hyperparameter we chose 0.0001. This architecture has 4 convolutional blocks, comprising a 2D convolutional layer with a 2D kernel (kernel size: 3×3), a 2D batch-normalization layer, a dropout rate (0.2), a rectified linear unit (ReLU) as an activation function and a max-pooling layer pool size=(2, 2). The number of feature channels for each block is 32, 64, 128, 256 respectively. The last three layers are fully connected layers to combine the feature vectors. In this study, the proposed 2D CNN model estimated 28,591,779 trainable parameters as shown in Table I.

TABLE I. 2D CNN MODEL FOR MRI.

| Layer (type) | Output Shape | Param # |
|---|---|---|
| conv2d_1 (Conv2D) | (None, 80, 80, 32) | 23072 |
| conv2d_2 (Conv2D) | (None, 80, 80, 64) | 18496 |
| conv2d_3 (Conv2D) | (None, 80, 80, 64) | 36928 |
| conv2d_4 (Conv2D) | (None, 80, 80, 64) | 36928 |
| conv2d_5 (Conv2D) | (None, 80, 80, 64) | 36928 |
| batch_normalization_1 (Batch | (None, 80, 80, 64) | 256 |
| max_pooling2d_1 (MaxPooling2 | (None, 40, 40, 64) | 0 |
| dropout_1 (Dropout) | (None, 40, 40, 64) | 0 |
| conv2d_6 (Conv2D) | (None, 40, 40, 128) | 73856 |
| conv2d_7 (Conv2D) | (None, 40, 40, 128) | 147584 |
| batch_normalization_2 (Batch | (None, 40, 40, 128) | 512 |
| max_pooling2d_2 (MaxPooling2 | (None, 20, 20, 128) | 0 |
| dropout_2 (Dropout) | (None, 20, 20, 128) | 0 |
| conv2d_8 (Conv2D) | (None, 20, 20, 256) | 295168 |
| conv2d_9 (Conv2D) | (None, 20, 20, 256) | 590080 |
| conv2d_10 (Conv2D) | (None, 20, 20, 256) | 590080 |
| batch_normalization_3 (Batch | (None, 20, 20, 256) | 1024 |
| max_pooling2d_3 (MaxPooling2 | (None, 10, 10, 256) | 0 |
| dropout_3 (Dropout) | (None, 10, 10, 256) | 0 |
| flatten_1 (Flatten) | (None, 25600) | 0 |
| dense_1 (Dense) | (None, 1024) | 26215424 |
| dropout_4 (Dropout) | (None, 1024) | 0 |
| dense_2 (Dense) | (None, 512) | 524800 |
| dropout_5 (Dropout) | (None, 512) | 0 |
| dense_3 (Dense) | (None, 3) | 1539 |

Total params: 28,592,675
Trainable params: 28,591,779
Non-trainable params: 896

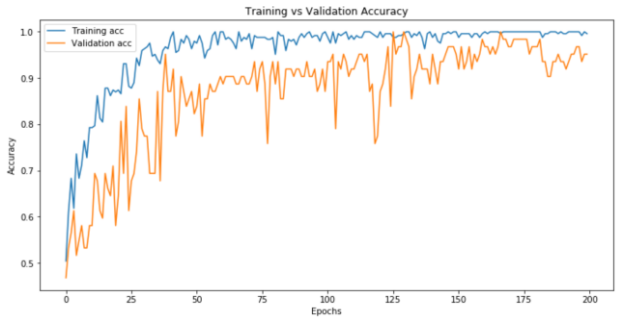

Fig. 5. Training and validation accuracy after 200 epochs for 2D model.

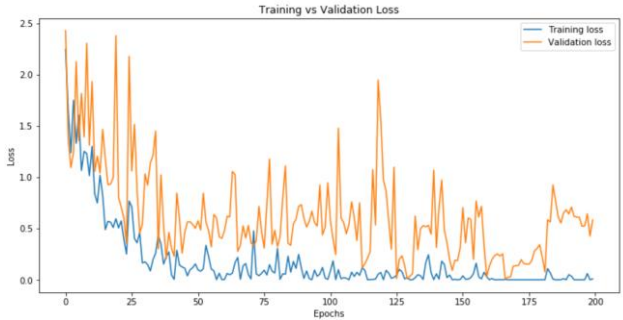

Fig. 6. Training and validation loss after 200 epochs for 2D model.

We use the term "Validation" in this study instead of "Test" because in medicine, "Validation" is used for the process of verifying the performance of the prediction model, which is equivalent to the term "Test" in machine learning.

TABLE II. RELATIONSHIP BETWEEN NEURODEVELOPMENTAL AGE AND BRAIN PREDICTED AGE IN A 2D CLASSIFICATION MODEL AFTER 200 EPOCHS.

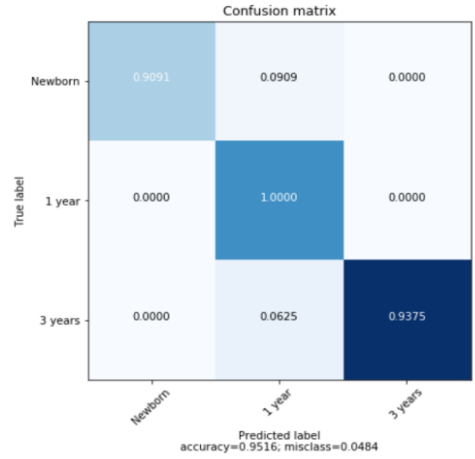

TABLE III. 2D CNN STATISTICAL RESULTS AFTER 200 EPOCHS.

| Class | Statistical result of 2D CNN after 200 epochs | | |
|---|---|---|---|
| | *Precision* | *Recall* | *F1-score* |
| newborn | 1.00 | 0.91 | 0.95 |
| 1 Year | 0.86 | 1.00 | 0.93 |
| 3 Years | 1.00 | 0.94 | 0.97 |

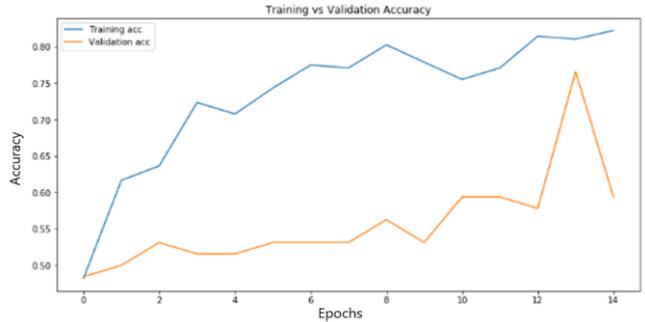

Fig. 7. Training and validation accuracy after 14 epochs for 2D model.

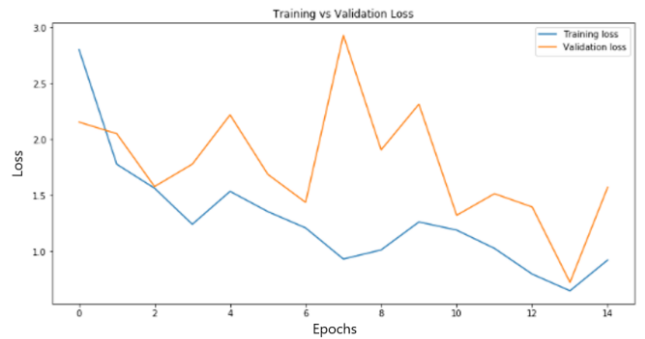

Fig. 8. Training and validation loss after 14 epochs for 2D model.

As seen in Fig.5 and 6, we achieved 95% accuracy after 200 epochs. Thus, 2D models need more epochs for training to obtain high accuracy. Tables II and III show the statistical analysis and normalized confusion matrix after 200 epochs. Fig. 7 and 8 depict accuracy and loss comparisons after 14 epochs in training and validation from the 2D model to compare with 3D model. After 14 epochs, we achieved 63% accuracy in training and 59% accuracy in validation.

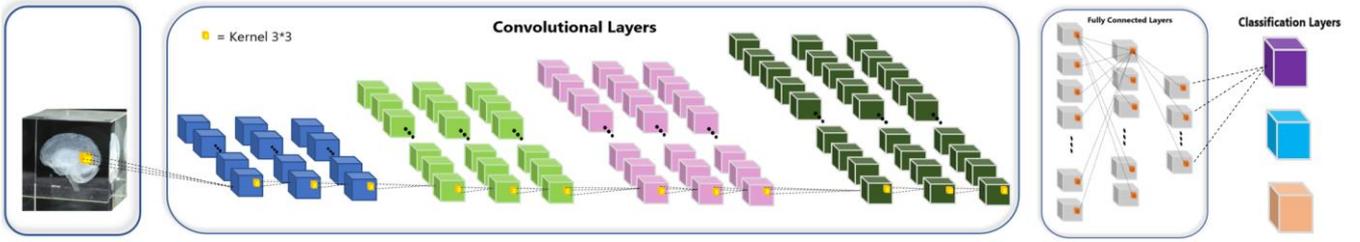

Fig. 9. The 3D convolutional neural network architecture. 3D boxes represent input and feature maps. The proposed neurodevelopmental age estimation architecture contains 4 blocks of 3D convolutional operation with 3D batch normalization ReLU and, max-pooling operation after each block of 3D convolutional layers. Three layers of a fully connected layer at the end, that generate the regression model to output predicted brain age in three age cohorts.

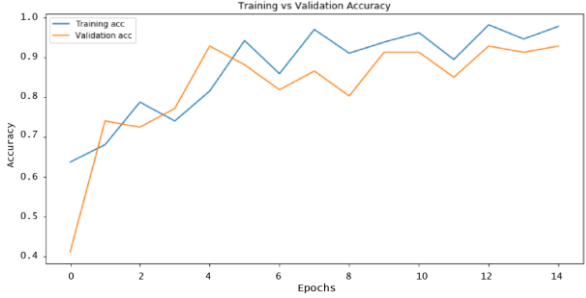

Fig. 10. Training and validation accuracy after 14 epochs for 3D model.

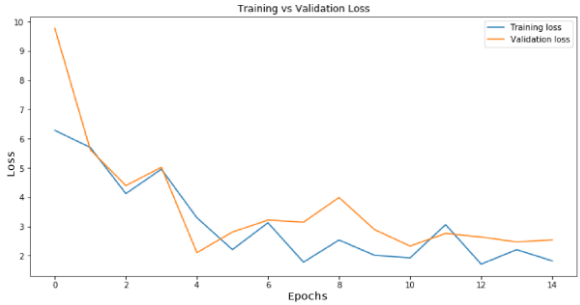

Fig. 11. Training and validation loss after 14 epochs for 3D model.

TABLE IV. 3D CNN MODEL FOR MRI.

```
Layer (type)                 Output Shape              Param #
=================================================================
conv3d_1 (Conv3D)            (None, 80, 80, 80, 32)    896
batch_normalization_1 (Batch (None, 80, 80, 80, 32)    128
max_pooling3d_1 (MaxPooling3 (None, 80, 40, 40, 32)    0
conv3d_2 (Conv3D)            (None, 78, 38, 38, 64)    55360
conv3d_3 (Conv3D)            (None, 76, 36, 36, 64)    110656
conv3d_4 (Conv3D)            (None, 74, 34, 34, 64)    110656
conv3d_5 (Conv3D)            (None, 72, 32, 32, 64)    110656
batch_normalization_2 (Batch (None, 72, 32, 32, 64)    256
max_pooling3d_2 (MaxPooling3 (None, 72, 16, 16, 64)    0
dropout_1 (Dropout)          (None, 72, 16, 16, 64)    0
conv3d_6 (Conv3D)            (None, 70, 14, 14, 128)   221312
conv3d_7 (Conv3D)            (None, 68, 12, 12, 128)   442496
batch_normalization_3 (Batch (None, 68, 12, 12, 128)   512
max_pooling3d_3 (MaxPooling3 (None, 68, 6, 6, 128)     0
dropout_2 (Dropout)          (None, 68, 6, 6, 128)     0
conv3d_8 (Conv3D)            (None, 67, 5, 5, 256)     262400
conv3d_9 (Conv3D)            (None, 66, 4, 4, 256)     524544
conv3d_10 (Conv3D)           (None, 64, 2, 2, 256)     1769728
batch_normalization_4 (Batch (None, 64, 2, 2, 256)     1024
max_pooling3d_4 (MaxPooling3 (None, 64, 1, 1, 256)     0
dropout_3 (Dropout)          (None, 64, 1, 1, 256)     0
flatten_1 (Flatten)          (None, 16384)             0
dense_1 (Dense)              (None, 1024)              16778240
dropout_4 (Dropout)          (None, 1024)              0
dense_2 (Dense)              (None, 512)               524800
dropout_5 (Dropout)          (None, 512)               0
dense_3 (Dense)              (None, 3)                 1539
=================================================================
Total params: 20,915,203
Trainable params: 20,914,243
Non-trainable params: 960
```

TABLE V. RELATIONSHIP BETWEEN NEURODEVELOPMENTAL AGE AND PREDICTED BRAIN AGE IN A 3D CLASSIFICATION MODEL AFTER 14 EPOCHS.

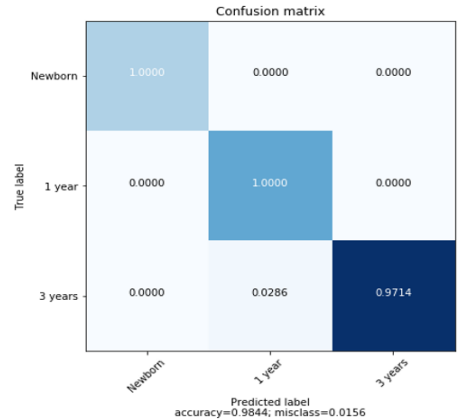

## B. 3D CNN Model

Figure 9 shows a 3D-CNN architecture used in the proposed neurodevelopmental age estimation methodology. As similar to 2D CNN, this architecture has 4 convolutional blocks consisting of a 3D convolutional layer (kernel size: 3×3×3), a 3D batch-normalization layer, dropout rate (0.2), ReLU activation layer and a max-pooling layer (pool_size=(1, 2, 2), strides=(1, 2, 2)). The number of feature channels is 32, 64, 128, and 256 for each block, respectively. The last three layers are fully connected layers to combine the feature vectors. The output is a scalar value of the estimated age of three age cohorts. we also used the same optimizer and learning rate as those in 2D CNN. These hyperparameters might be the best for our proposed 3D CNN as shown in Table IV. However, we tried various optimizer, kernel size, and learning rate. Fig. 10 and 11 compare the accuracy and loss function after 14 epochs in training and validation. The proposed 3D CNN model required 150 seconds of training to estimate 20,914,243 trainable parameters with the GPU for 14 epochs. Table V shows a normalized confusion matrix of our proposed 3D CNN in each class. In this table, the percentage of numbers is correct in each class, and the numbers of other cells show the error of the BDAE system. With this model, we achieved 100% accuracy in training 98.4% accuracy in validation after 14 epochs.

TABLE VI. 3D CNN STATISTICAL RESULTS AFTER 14 EPOCHS.

| Class | Statistical Result of 3D CNN | | |
|---|---|---|---|
| | *Precision* | *Recall* | *F1-score* |
| newborn | 1.00 | 1.00 | 1.00 |
| 1 Year | 0.95 | 1.00 | 0.97 |
| 3 Years | 1.00 | 0.97 | 0.99 |

As demonstrated by the developmental brain results of 3D CNN in Table VI and compared with those of 2D CNN, we need more than 200 epochs to obtain the same accuracy of 3D CNN after 14 epochs. Because 2D CNNs take a single slice as input, therefore, we do not have the information of adjacent slices. Thus, losing the information of regions of interest in the brain is more likely. Having voxel information from adjacent slices would be useful for the estimation in multiclassification. According to the statistical results of the two models, the 3D CNN model addresses this issue by using 3D convolutional kernels for classification. This ability of 3D CNN can lead to improved performance. Therefore, when volumetric content is important in data analysis, 3D CNN is preferred. The faster classification provided by 3D CNN also allows doctors to expedite patient care, especially to prevent long-term complications in infected infants. Beyond infection, the 3D healthy brain model is an important tool that could aid the radiologist and referring clinicians in identifying abnormal brain structure which could be due to any number of causes. At that point, further testing could be done to confirm underlying infectious, genetic, or traumatic etiologies.

## VI. CONCLUSION

To our knowledge, this study is the first to demonstrate that 3D CNNs can be used to accurately estimate neurodevelopmental age in infants based on brain MRIs. The methods discussed here accurately classified neurodevelopmental age from T1, T2, and PD MRI brain sequences of healthy infants from the NIHPD dataset. This 98.4% accuracy was achieved using a common MRI sequence dataset, with minimal processing necessary to generate an accurate neurodevelopmental age estimation. Compared to 2D CNN, 3D CNNs had greater accuracy and faster processing. Our success in accurately classifying MRIs into the 3 difficult and developmentally important age groups of the newborn, 1 year and 3 years, allows us to refine these methods to define normative MRI patterns in smaller childhood age brackets. Our result demonstrated the feasibility of accurately estimating normative structural brain development using 3D CNN to estimate neurodevelopmental age in infants in the first years of life. These techniques will be useful to assess structural brain neurodevelopment in children suspected of having conditions potentially affecting brain development and predict their outcomes.

ACKNOWLEDGMENT

We would like to express our special thanks to the NIMH Data Archive for access to the NIH pediatric MRI dataset.